\documentclass[%
 reprint,
superscriptaddress,
$preprintnumbers,
 amsmath,amssymb,
 aps,
]{revtex4-1}

\usepackage{graphicx}
\usepackage{dcolumn}
\usepackage{bm}

\begin{document}

\preprint{APS/123-QED}

\title{Using physical features of protein core packing to distinguish real proteins from decoys}

\author{Alex T. Grigas}
 \affiliation{Graduate Program in Computational Biology and Bioinformatics, Yale University, New Haven, Connecticut, 06520, USA}
 \affiliation{Integrated Graduate Program in Physical and Engineering Biology, Yale University, New Haven, Connecticut, 06520, USA}
\author{Zhe Mei}
\affiliation{Integrated Graduate Program in Physical and Engineering Biology, Yale University, New Haven, Connecticut, 06520, USA}
\affiliation{Department of Chemistry, Yale University, New Haven, Connecticut 06520, USA}
\author{John D. Treado}
\affiliation{Integrated Graduate Program in Physical and Engineering Biology, Yale University, New Haven, Connecticut, 06520, USA}
\affiliation{Department of Mechanical Engineering and Materials Science, Yale University, New Haven, Connecticut 06520, USA}
\author{Zachary A. Levine}
\affiliation{Department of Pathology, Yale University, New Haven, Connecticut 06520, USA}
\affiliation{Department of Molecular Biophysics and Biochemistry, Yale University, New Haven, Connecticut, 06520}
\author{Lynne Regan}
\affiliation{Institute of Quantitative Biology, Biochemistry and Biotechnology, Centre for Synthetic and Systems Biology, School of Biological Sciences, University of Edinburgh}
\author{Corey S. O'Hern}
 \email{corey.ohern@yale.edu}
\affiliation{Graduate Program in Computational Biology and Bioinformatics, Yale University, New Haven, Connecticut, 06520, USA}
\affiliation{Integrated Graduate Program in Physical and Engineering Biology, Yale University, New Haven, Connecticut, 06520, USA}
\affiliation{Department of Mechanical Engineering and Materials Science, Yale University, New Haven, Connecticut 06520, USA}
\affiliation{Department of Physics, Yale University, New Haven, Connecticut 06520, USA}
\affiliation{Department of Applied Physics, Yale University, New Haven, Connecticut 06520, USA}

\begin{abstract}
The ability to consistently distinguish real protein structures from
computationally generated model decoys is not yet a solved problem. One route to distinguish
real protein structures from decoys is to delineate the
important physical features that specify a real protein. For example,
it has long been appreciated that the hydrophobic cores of proteins
contribute significantly to their stability.  As a dataset of
decoys to compare with real protein structures, we studied submissions
to the bi-annual CASP competition (specifically CASP11, 12, and 13),
in which researchers attempt to predict the structure of a protein
only knowing its amino acid sequence.  Our analysis reveals that many
of the submissions possess cores that do not recapitulate the features
that define real proteins. In particular, the model structures appear
more densely packed (because of energetically unfavorable atomic
overlaps), contain too few residues in the core, and have improper
distributions of hydrophobic residues throughout the structure.  Based
on these observations, we developed a deep learning method, which
incorporates key physical features of protein cores, to
predict how well a computational model recapitulates the real protein
structure without knowledge of the structure of the target
sequence. By identifying the important features of protein structure,
our method is able to rank decoys from the CASP competitions
equally well, if not better than, state-of-the-art methods that
incorporate many additional features.
\end{abstract}


\maketitle

\section{\label{sec:intro}Introduction}
\begin{figure*}
\centering
\includegraphics[width=.8\textwidth]{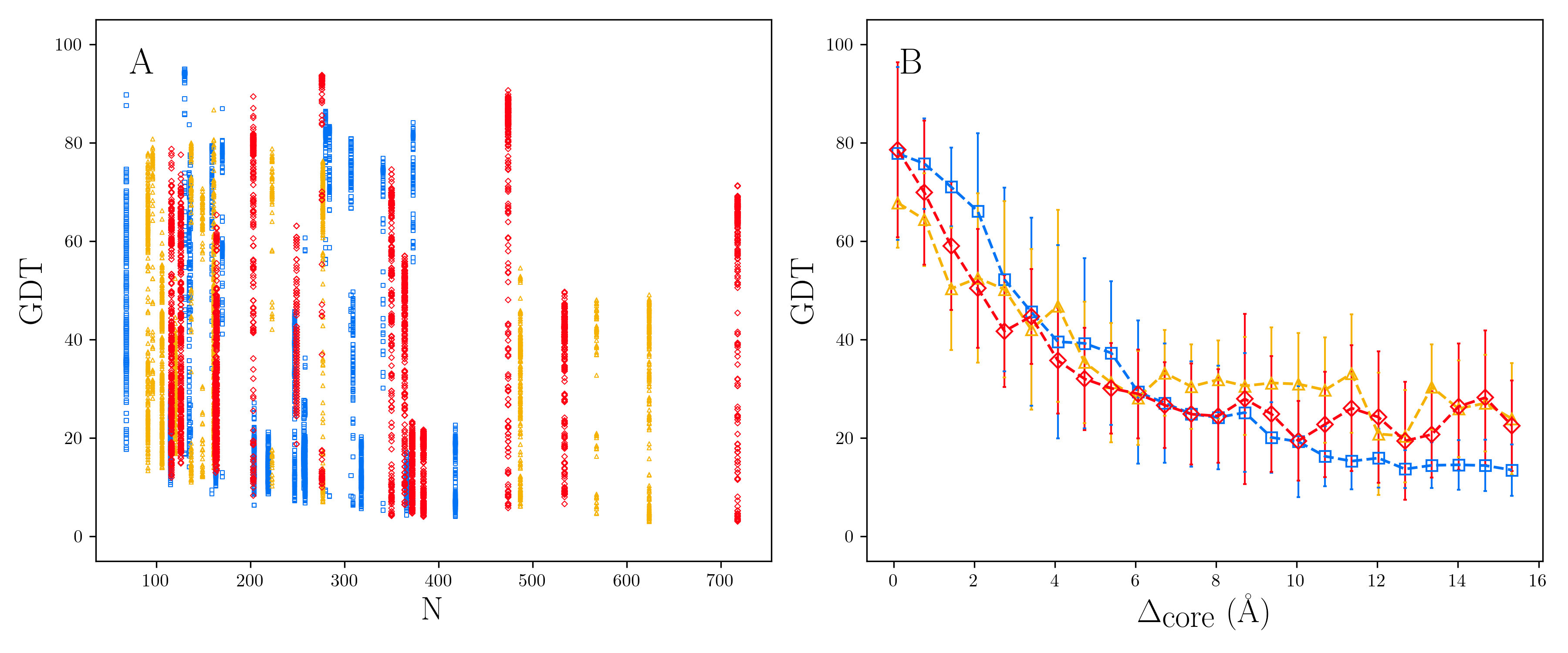}
\caption{(A) Scatter plot of the Global Distance Test (GDT) score, which gives the average
percentage of $C_{\alpha}$ atoms that is within a given cutoff
distance to the target (averaged over four cutoff distances), versus the number of residues $N$ in the target structure
for free modeling submissions to CASP11 (blue squares), CASP12
(orange triangles), and CASP13 (red diamonds). (B) GDT plotted
versus the root-mean-square deviations (RMSD) among C$_{\alpha}$
atoms of core residues defined in the target ($\Delta_{\rm
core}$). The symbols represent the average in each $\Delta_{\rm
core}$ bin and the error bars represent one standard deviation.}
\label{fig:gdt_size}
\end{figure*}

It remains a grand challenge of biology to design proteins
that adopt user-specified structures and perform user-specified
functions. Although there have been significant
successes~\citep{designsuccess:BakerScience2003,designsuccess:ButterfosAnnRev2006,designsuccess:YinScience2007,designsuccess:JiangScience2008,design:RocklinScience2017,design:DeGradoScience1988,design:Bryson1995,design:LanciPNAS2012,design:ThomsonScience2014,design:DawsonCurrOpinChemBio2019,lynneref:MainStruc2003},
the field is still not at the point where we can robustly achieve this
goal for any application~\citep{designreview:BakerProtScience2019}.
An inherent problem in protein structure prediction and design is that
it is extremely difficult to distinguish between computational models
that are apparently low
energy~\citep{decoyreview:ZhangCurrOpinStrucBio}, but which are
different from the real, experimentally determined
structures~\citep{CASP:KryshtafovychProteins2019,deshaw:RobustelliPNAS2018,deshaw:Lindorff-Larsen2011}. This
problem is known as ``Decoy Detection''.  For example, in recent
Critical Assessment of protein Structure Prediction (CASP)
competitions, in which researchers attempt to predict the three-dimensional 
(3D) structure
of a protein, based on its amino acid sequence, many groups produced
impressively accurate predictions for certain targets (Fig.~\ref{fig:gdt_size}
(A)). However, for most targets there is a wide spread of prediction
accuracy across the submissions from different groups. (Note that the fluctuations in prediction accuracy across groups is comparable to fluctuations 
within a single group. See Supplementary Information (SI).)

In recognition of this issue, there is a subcategory in CASP,
Estimation of Model Accuracy (EMA), in which researchers aim to rank
order the submitted models according to their similarity to the
backbone of the target structure. The challenge is that researchers 
must develop such a scoring function for determining model accuracy, yet they do not have access to the
target
structure~\citep{ema:CozzettoProteins2007,ema:CozzettoProteins2009,ema:KryshtafovychProteins2011,ema:KryshtafovychProteins2014,ema:KyshtafovychProteins2016,ema:KryshtafovychProteins2017,ema:ChengProteins2019}. Although EMA methods are
improving~\citep{DOPE:ShenProteinscience2006,RW:ZhangPLoS2010,OPUS:LuJMolBio2008,SBROD:Karasikovbioinformatics2018,ProQ2:RayBioinformatics2012,ProQ3:UzielaScientificReports2016,QMean:BenkertBioinformatics2010,QMeanDisCo:WaterhouseNucleicAcids2018,GOAP:ZhouBPJ2011,DFIRE:ZhouProteinScience2009,VoroMQA:OlechnoviProteins2017},
they are still unable to consistently rank models submitted to CASP in
terms of their similarity to the target
structure~\citep{ema:ChengProteins2019}.

The protein core has long been known to determine protein stability
and provide the driving force for
folding~\citep{folding:DillBiochemistry1990,core:RichardsQRevBioPhys1993,lynneref:MunsonProtStruc1994,lynneref:MunsonProtStruc1996,lynneref:WillisStruc2000,lynneref:DalalFold1997,lynneref:DalalProtSci2000,lynneref:ReganPepSci2015,lynneref:RichardsAnnRev1977}. Additionally,
in our previous work, we have found that several features of core
packing are universal among well-folded experimental structures, such
as the repacking predictability of core residue side chain placement,
core packing fraction, and distribution of core void
space~\citep{subgroup:GainesPRE2016,subgroup:TreadoPRE2019,subgroup:GainesProteins2018,subgroup:CaballeroPEDS2016,subgroup:GainesPED2017,subgroup:GainesJoPCM2017}. This
work suggests that analysis of core residue placement and packing 
in proteins more generally should be a
powerful tool for determining the accuracy of protein decoys. Indeed,
the RosettaHoles software uses defects in interior void space to
differentiate between high-resolution x-ray crystal structures and protein
decoys~\citep{rosettaholes:ShefflerProteinSci2009}. Nevertheless, a
minimal set of features that can determine protein decoy accuracy has
not yet been identified.

We demonstrate, that for recent CASP competition predictions, we can
determine protein decoy accuracy solely by identifying the structures
that place the correct residues in the protein core. We also show that
only predicted structures that place core residues accurately,
measured using the root-mean-squared deviation of the C$_{\alpha}$ atoms of
solvent inaccessible residues (i.e. $\Delta_{\rm core} < 1$\AA), can
achieve high Global Distance Test (GDT) scores (GDT $\gtrsim 70$)
(Fig.~\ref{fig:gdt_size} (B)), where GDT ranges from $0$ to $100$ and $100$ is a perfect match to the target structure~\citep{GDT:ZemlaNuclieicAcids2003}. Motivated by these observations, we then
analyzed several important attributes of the {\it cores} of both
experimentally-observed and predicted protein structures. Using these
results, we developed a decoy detection method based on only five
principal features of protein packing that are independent of the target
structure. Our method is more effective than many of the methods
in the CASP13 EMA. Moreover, all of the methods used in CASP13 EMA employ a far greater number of
features than we do~\citep{ema:WonProteins2019}. For example, in contrast to our
approach, the top performing method in the CASP13 EMA,
ModFOLD7~\citep{ema:ChengProteins2019,ema:WonProteins2019}, uses a neural 
network to combine
$21$ scoring metrics, each based on numerous starting features, 
to reach a ``consensus'' GDT.  The
effectiveness of the small number of features in our approach
highlights the importance of core residues, which take up $\lesssim 10$\% of
globular proteins on average, and packing constraints in determining the global structure of
proteins.

\section{Results}

First, we identify several key features that distinguish
high-resolution x-ray crystal structures and computationally-generated
decoys, such as the average core packing fraction, core overlap
energy, fraction of residues positioned in the core, and the
distribution of the packing fraction of hydrophobic residues throughout
the protein. We then show how these features can be used to predict
the GDT of CASP submissions, independent of knowing the target
structure.

\begin{figure*}
\centering
\includegraphics[width=.75\textwidth]{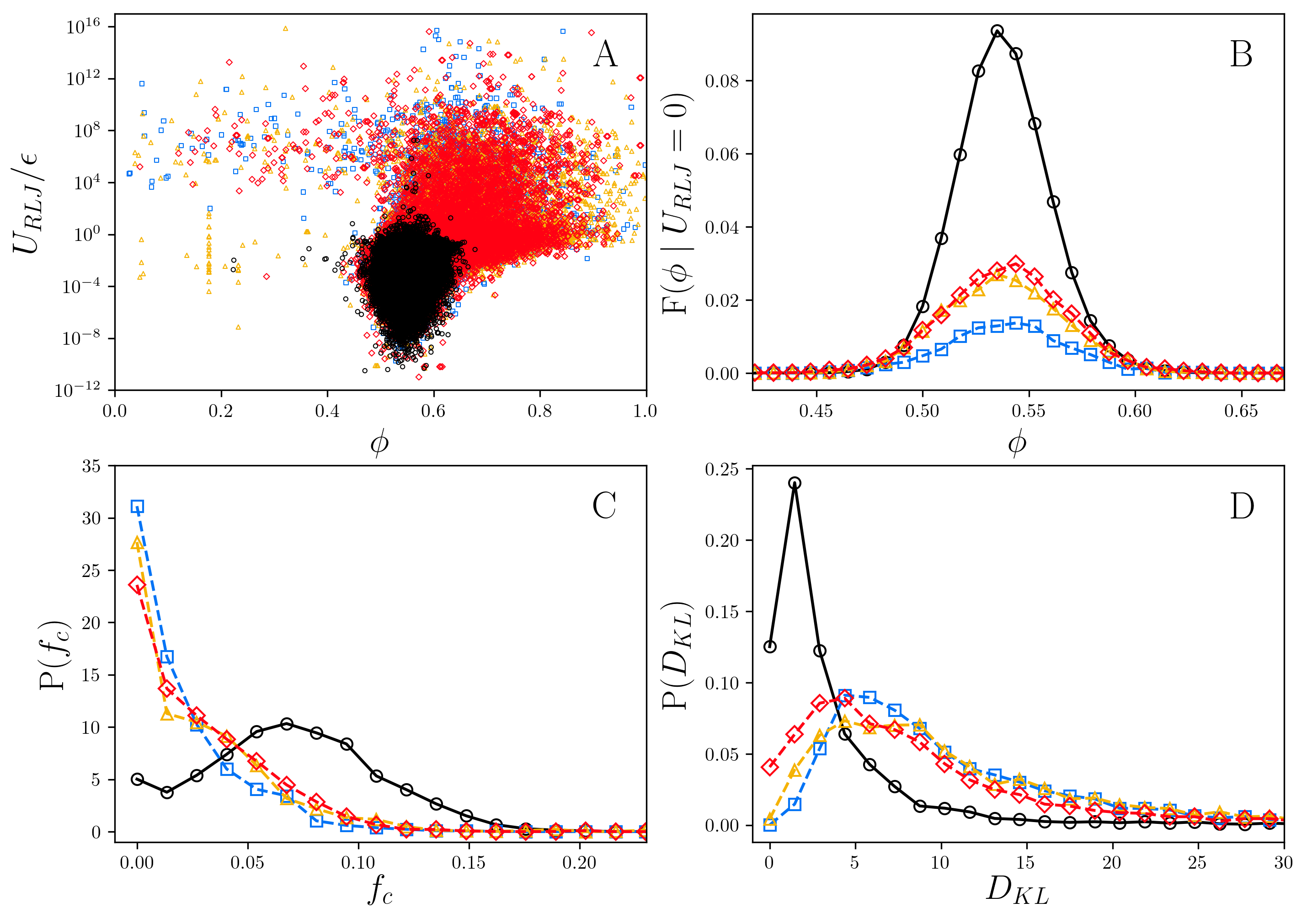}
\caption{Packing features of high-resolution x-ray crystal 
structures (black circles) and submissions
to CASP11 (blue squares), CASP12 (orange triangles), and CASP13 (red
diamonds). (A) Purely repulsive Lennard-Jones potential energy $U_{\rm RLJ}$ 
that measures the overlap of core
residue sidechain atoms versus packing fraction $\phi$. (B) Frequency
distribution of the packing fraction $F(\phi | U_{\rm RLJ}=0)$ for core 
residues with zero overlap
energy. (C) Probability distribution $P(f_{c})$ of the fraction of core
residues $f_{c}$. (D) Probability distribution $P(D_{KL})$ of the 
Kullback-Leibler divergence $D_{KL}$ from the distribution of
the packing fractions of all hydrophobic residues in high-resolution 
x-ray crystal structures.}
\label{fig:features}
\end{figure*}

The distribution of packing fractions $\phi$ of core residues in
proteins whose structures are determined by x-ray crystallography
occur over a relatively narrow range, with a mean of $0.55$ and a
standard deviation of
$0.1$~\citep{subgroup:GainesPRE2016,subgroup:GainesProteins2018,subgroup:GainesJoPCM2017}. We define core
residues as those with small values of the relative
solvent accessible surface area, ${\rm rSASA} < 10^{-3}$.  (See the
Materials and Methods section for a description of the database of high-resolution protein 
x-ray crystal structures and definition of rSASA.) In contrast, we find that many of the CASP
submissions possess core residues with packing fractions that are much
higher than those in experimentally determined proteins structures. One way to achieve such an un-physically
high packing fraction would be to allow atomic overlaps. We therefore
analyzed the side-chain overlap energy for core residues, using the
purely repulsive Lennard-Jones inter-atomic potential,
\begin{equation}
U_{\rm RLJ} = N_a^{-1} \sum_{i,j}\frac{\epsilon}{72}\left(1-\left(\frac{\sigma_{ij}}{r_{ij}}\right)^6\right)^2\Theta(\sigma_{ij}-r_{ij}),
\end{equation}
where the sum is taken over all side-chain atoms $i$ and all other
atoms not part of the same residue $j$, $\epsilon$ defines the energy
scale, $\sigma_{ij}=(\sigma_{i}+\sigma_{j})/2$, $\sigma_{i}$ is
the diameter of atom $i$, $r_{ij}$ is the distance between atoms $i$
and $j$, and $\Theta(x)$ is the Heaviside step function, which is 1
when $x>0$ and is $0$ when $x\leq0$. For high-resolution x-ray crystal
structures, half of core residues have an overlap energy of zero;
the remaining half of the residues have very small overlap energies with an average
value of $U_{\rm RLJ}/\epsilon \approx 10^{-4}$
(Figs.~\ref{fig:features} (A) and (B)). In contrast, the models in the CASP datasets
include some extremely high energy residues, with $U_{\rm
  RLJ}/\epsilon \sim 10^{16}$. The absence of data points in
the lower right-hand corner of Fig.~\ref{fig:features} (A)
clearly highlights that artificially high packing fractions are only found
when the overlap energy is high. In Fig.~\ref{fig:features} (B), we
show the frequency distribution of packing fractions for core
residues with $U_{\rm RLJ} = 0$. The differences in peak heights
reflect how much more likely it is for core
residues from x-ray crystal structures of proteins to have zero overlap 
energy compared to those in the CASP
submissions.

These results demonstrate that individual core residues in
the computational models submitted to CASP are typically overpacked. We then asked whether core
overpacking is related to the number of residues in the core relative
to the number of residues in the protein. In Fig.~\ref{fig:features} (C), we plot the
probability that a structure, either computationally-generated or
experimentally-determined, has a given fraction of its total number of
residues in the core. It is clear from this plot that
computationally-generated models often have too few residues in the
core. Thus, the computationally-generated models not only possess
cores with un-physically high packing fraction and overlap energy,
but they also, typically, have a smaller fraction of residues in the
core compared to x-ray crystal structures of proteins.

Many CASP models have too few residues in the core; how does this
affect the distribution of hydrophobic residues outside of the core?
We examined the degree to which the packing fractions of all
hydrophobic residues in a given protein deviate from the expected
distribution from high-resolution x-ray crystal
structures~\citep{PISCES:WangBioinformatics2003,PISCES:WangNucleicAcids2005}. (See
Fig.~\ref{fig:features} (D).) Specifically, we measured the
Kullback-Leibler (KL) divergence ($D_{KL}$) between the overall
distribution of packing fractions of hydrophobic residues from a
database of high-resolution x-ray crystal structures, and each
individual structure's packing fraction distribution for all its
hydrophobic residues in that
database~\citep{kl_div:KullbackAnnalMathStat}. (See SI for more
details.) Additionally, we measured the $D_{KL}$ for all CASP models
against the distribution from the database of high-resolution x-ray
crystal structures. We find that the distribution of packing fractions
of hydrophobic residues for each individual experimentally-observed
protein structure is similar to the full distribution, whereas the
distributions for the computationally-generated structures differ
significantly from the experimentally observed distribution.

\begin{figure}[b]
\centering
\includegraphics[width=.42\textwidth]{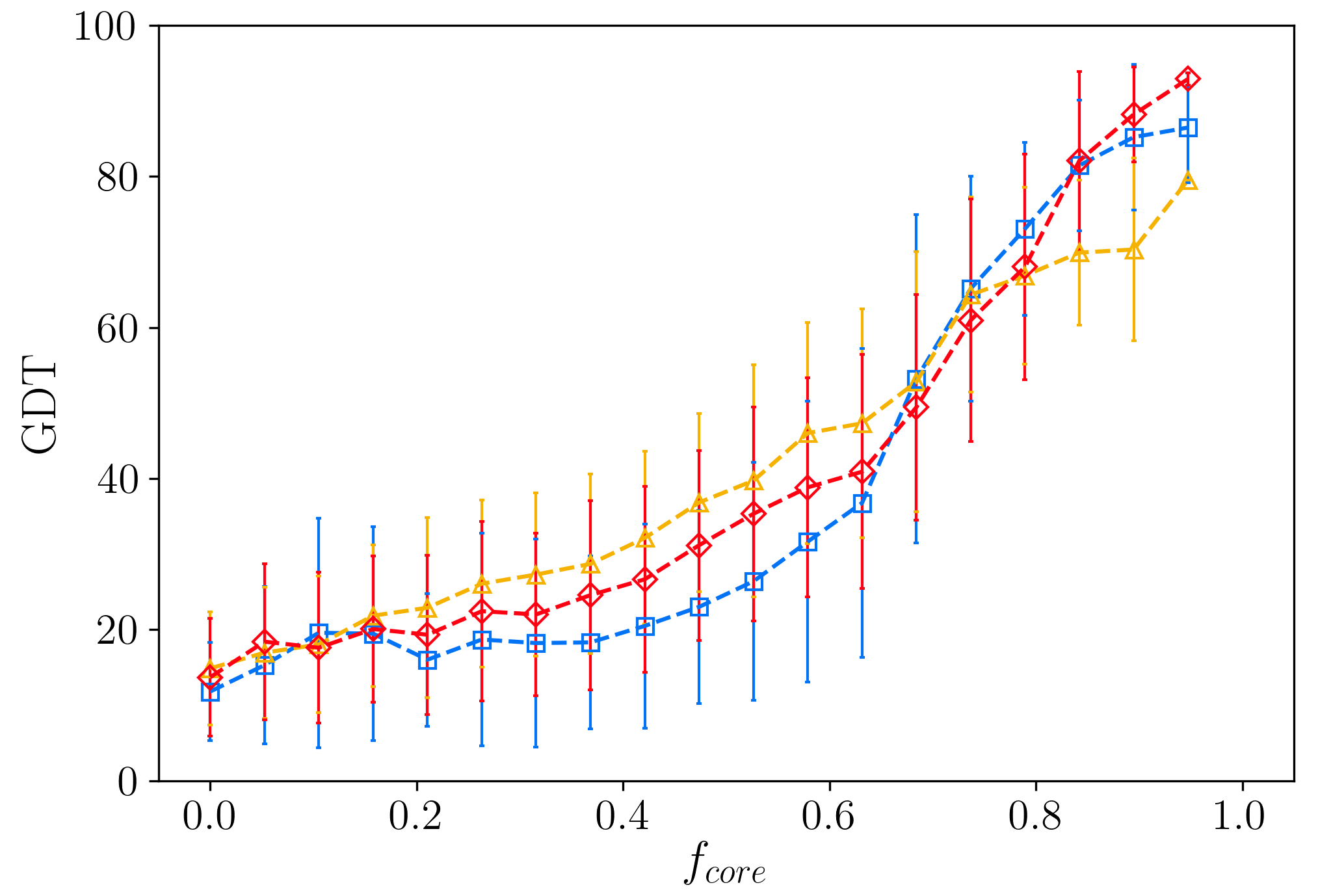}
\caption{The average GDT of CASP predictions that correctly identify each 
given fraction of near core residues with ${\rm rSASA} \leq 10^{-1}$, $f_{\rm core}$, for CASP11 (blue squares), CASP12 (orange triangles), and CASP13 (red diamonds) structures. Error bars represent one standard deviation.}
\label{fig:core_identity}
\end{figure}

Before developing a predictive model for decoy detection, we
investigated the correlation between the accuracy of backbone
placement and correct identification of core residues.  In
Fig.~\ref{fig:core_identity}, we plot the average GDT versus the
fraction $f_{\rm core}$ of the predicted core residues that are core
residues in the target structure. This plot shows that there is a
strong correlation between the accuracy of backbone placement and
correct identification of the core residues. In particular, when
$f_{\rm core} \rightarrow 1$, the average ${\rm GDT} \gtrsim
80$. However, one does not know the correct set of core residues at the
time of the prediction. Yet, the core residues should share the features
shown in Fig.~\ref{fig:features}. Therefore, we should be able to predict the GDT of a model based upon how well the core properties and the distribution of the hydrophobic residues match those of high-resolution x-ray crystal structures of proteins.

\begin{figure}
\centering
\includegraphics[width=.45\textwidth]{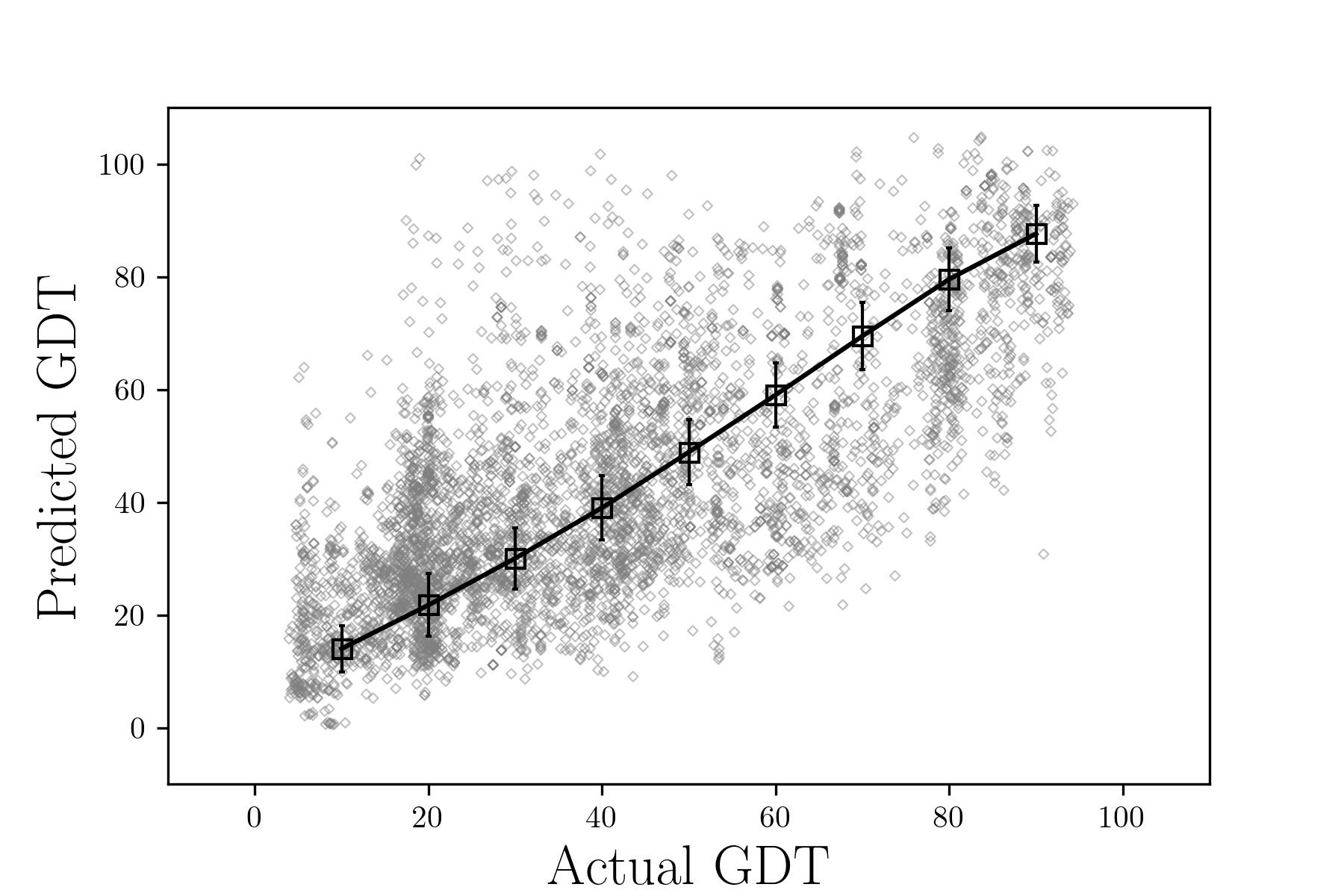}
\caption{Predicted versus actual GDT of CASP13 structures (gray diamonds) from a 
model that was developed from the four 
features in Fig.~\ref{fig:features} plus $N$ input into a neural network. The open squares represent the average
value of the predicted GDT in each GDT bin and the error bars represent
one standard devation.}
\label{fig:machine_learning}
\end{figure}

While we have shown that many predicted structures submitted to CASP do not
recapitulate the packing properties of high-resolution protein x-ray crystal
structures, we have not yet made a quantitative link between
differences in these properties and the overall backbone accuracy
(i.e. GDT). Therefore, we developed a neural network based on the four
packing-related features in Fig.~\ref{fig:features}, plus the number,
$N$, of residues in the protein, to construct the GDT function. (We
included $N$ to account for larger fluctuations in packing properties
that occur for small $N$.) We built a simple feed-forward neural
network with five hidden layers and a combination of common non-linear
activation functions. (For more details, see SI.) The mean-squared error in GDT was used as the loss
function. Submissions from CASP11, CASP12, and a large database of
high-resolution x-ray crystal structures~\citep{PISCES:WangBioinformatics2003,PISCES:WangNucleicAcids2005} were used as training
data. The model was then tested on CASP13 submissions. The results for
the predicted versus actual GDT are plotted in
Fig.~\ref{fig:machine_learning}. Our model achieves a Pearson
correlation of $0.72$, a Spearman correlation of $0.71$, a Kendall Tau
of $0.51$, and an average absolute error of $13$ GDT. For comparison,
in the most recent assessment of decoy detection (EMA 13), one of the
top ranked single-ended methods, ProQ3, reported a correlation between
CASP13 actual GDT and predicted GDT of
$0.67$~\citep{ema:ChengProteins2019}. Another recent study reported a
maximum Pearson correlation of $0.66$ for predicted versus actual GDT
for several methods that tested on CASP12
structures~\citep{SBROD:Karasikovbioinformatics2018}. The best
absolute GDT loss reported in the CASP13 EMA competition was $7$ GDT
and the average GDT loss across all methods was
$15$~\citep{ema:WonProteins2019}.

\begin{figure}
\centering
\includegraphics[width=.45\textwidth]{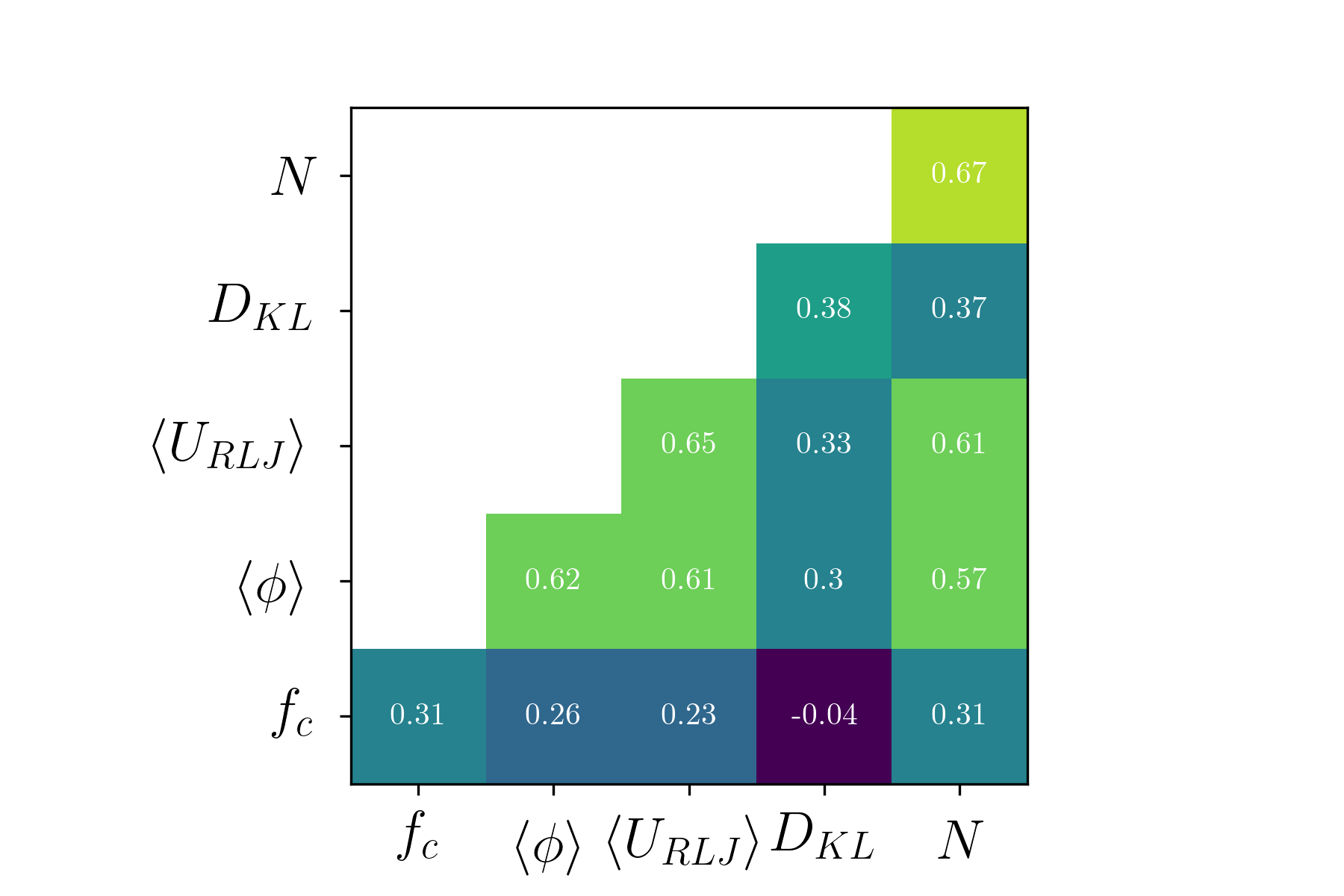}
\caption{Pearson correlation coefficients between the predicted and 
actual GDT of CASP13 structures following permutations of single features (along the diagonal) and pairs of features (for the off-diagonal 
components). The color ranges from purple (0) to yellow (1) corresponding to the Pearson correlation coefficient.}
\label{fig:permutation}
\end{figure}

We also investigated the importance of each feature in the neural
network model. To do this, we randomly permuted the values of a given
feature after training. This procedure decorrelates each structure with
its feature value to effectively remove that feature from the
model. In Fig.~\ref{fig:permutation}, we display the Pearson
correlation between the predicted and actual GDT following feature
permutations, averaged over $200$ different random
permutations. All of the features are important, although
eliminating the sequence length, $N$, as a feature still yields a Pearson
correlation of $0.65$, indicating it is the least important. The two largest single feature changes come
from permuting either the fraction of core residues or the KL divergence
from the hydrophobic residue packing fraction distribution, leading to
Pearson correlations of $0.42$ and $0.39$, respectively. Also,
permuting both of these features together leads to the largest
pair-wise drop in the Pearson correlation to $\approx 0$.  These
results indicate that the most important pair of features to include
in protein decoy detection are the fraction of core residues and
packing fraction distribution of hydrophobic residues.  The packing 
fraction and overlap energy of core residues are slightly less 
important features. We believe this is because including the wrong residue in the core
will give rise to a low GDT (Fig.~\ref{fig:core_identity}), even if the packing fraction and overlap 
energy of the misplaced residues are typical of those for core residues in 
high-resolution protein x-ray crystal structures.

\section{Discussion}

We have identified several important features characterizing protein
packing that allow us to distinguish protein decoys from
experimentally realizable structures. We developed a machine learning
model, using deep learning on a small number of packing features, that is able to predict
the GDT of CASP13 structures with high accuracy and without knowledge
of the target structures. In addition to developing a highly
predictive model, this work also demonstrates the importance of the
core and packing constraints for protein structure prediction and
points out potential improvements to current prediction methods
by properly modeling protein cores. Importantly, the machine learning model we
developed can be used to identify protein decoys beyond those
generated by CASP. For example, molecular dynamics (MD) simulations
are often used to analyze thermal fluctuations in folded proteins. To what
extent do the protein conformations sampled in such MD simulations
recapitulate the packing properties of experimentally observed protein structures~\citep{subgroup:MeiProteins2019}? The model developed
here can be used in concert with MD simulations to filter out
un-physical conformations, which will have low values of GDT, without using
knowledge of the experimentally observed protein structure. Thus, such
an approach can be used to improve protein structure
prediction. Additionally, our model can be used to assist protein
design methods by selecting designs that are more likely to be
experimentally attainable.

We expect future improvements to our basic model will increase its accuracy. For example, we
have shown that the identification of core residues is one of the most
important aspects for determining a predicted structure's
accuracy. Thus, we will also implement recurrent neural networks to
predict the ${\rm rSASA}$ values for each
residue~\citep{rsasa:HeffernanSciRep2015}. This model can then be
concatenated with the model developed here. In addition, we will
incorporate predictions of GDT into MD folding simulations to improve
the accuracy of computationally-generated protein structures. In addition to appreciating the overall success of our approach, it will also be informative to study in greater depth cases where there are large deviations in GDT. For example, investigating examples of high predicted GDT, but low actual GDT (or {\it vice versa}) has the potential to provide key insights into native protein structures.

\section{Materials and Methods}

\subsection*{Datasets}

In the main text, we show results for the free modeling CASP
submissions, and the corresponding results for template-based modeling
data are provided in the Supplementary Information. For the decoy
datasets, we examined CASP11 (2014)~\citep{CASP:MoultProteins2016},
CASP12 (2016)~\citep{CASP:MoultProteins2018} and CASP13 (2018)~\citep{CASP:KryshtafovychProteins2019} downloaded from the {\tt
  predictioncenter.org} data archive. Each target in the competitions
has a corresponding experimental structure. We selected targets with
an x-ray crystal structure under a resolution cutoff. A cutoff of $\leq 2.0$~\AA~was used in the
cases of CASP11 and CASP12, however; a cutoff of $\leq 2.7$~\AA~was
used for CASP13, as very few protein targets fell under $\leq
2.0$~\AA~. These cutoffs resulted in a dataset of $16,905$ predictions
based on $49$ target structures. For the x-ray crystal structure
dataset, we compiled a dataset of $5547$ x-ray crystal structures
culled from the PDB using
PISCES~\citep{PISCES:WangBioinformatics2003,PISCES:WangNucleicAcids2005}
with resolution $\leq 1.8$~\AA, a sequence identity cutoff of 20\%,
and an R-factor cutoff of $0.25$. 

\subsection*{rSASA}

To identify core residues, we measured each
residue's solvent accessible surface area (SASA). To calculate SASA,
we use the \textsc{Naccess} software
package~\citep{naccess:Hubbard1993}, which implements an algorithm
originally proposed by Lee and Richards~\citep{rsasa:LeeJMB1971}. To normalize the
SASA, we take the ratio of the SASA within the context of
the protein ($\text{SASA}_{\text{context}}$) and the SASA of the same
residue extracted from the protein structure as a dipeptide
(Gly-X-Gly) with the same backbone and side-chain dihedral angles:
\begin{equation}
\text{rSASA} = \frac{\text{SASA}_{\text{context}}}{\text{SASA}_{\text{dipeptide}}}.
\end{equation}
Core residues are classified as those that have ${\rm rSASA} \leq
10^{-3}$.  In Fig.~\ref{fig:core_identity}, ``near core'' residues are
those with ${\rm rSASA} \leq 10^{-1}$.

\subsection*{Packing Fraction}

A characteristic measure of the packing efficiency of a system is 
the packing fraction.  The packing fraction of residue $\mu$ is 
\begin{equation}
\phi_{\mu} = \frac{\nu_{\mu}}{V_{\mu}},
\end{equation}
where $\nu_{\mu}$ is the non-overlapping volume and $V_{\mu}$
is the volume of the Voronoi cell surrounding residue $\mu$. The Voronoi cell
represents the local free space around the residue. To
calculate the Voronoi tessellation for a protein structure, we use the
surface Voronoi tessellation, which defines a Voronoi cell as the
region of space in a given system that is closer to the bounding
surface of the residue than to the bounding surface of any other
residue in the system. We calculate the surface Voronoi tessellations
using the Pomelo software package~\citep{pomelo:WeisEPJ2017}. This
software approximates the bounding surfaces of each residue by
triangulating points on the residue surfaces. We find that using $\sim 400$
points per atom, or $\sim 6400$ surface points per residue, gives an
accurate representation of the Voronoi cells and the results
do not change if more surface points are included.



\bibliography{pnas-sample}

\end{document}